\documentclass[useAMS,usenatbib]{mn2e}
\usepackage[dvips]{graphicx}
\usepackage{amsmath}
\usepackage{hyperref}
\usepackage{xcolor}
\newcommand{\bmodulus}{$\langle B \rangle$}
\newcommand{\bz}{$\langle B_z \rangle$}

\newcommand{\vsini}{$v \sin i$}
\newcommand{\kms}{km\,s$^{-1}$}

\newcommand{\rsun}{R$_\odot$}
\newcommand{\msun}{M$_\odot$}

\newcommand{\teff}{$T_{\rm eff}$}

\newcommand{\rk}{$R_{\rm K}$}

\newcommand{\mnras}{$MNRAS$}
\newcommand{\apj}{$ApJ$}
\newcommand{\apjl}{$ApJL$}
\newcommand{\aj}{$AJ$}
\newcommand{\aap}{A\&A}

\newcommand{\aaps}{A\&AS}

\newcommand{\pasp}{PASP}
\newcommand{\apjs}{ApJS}
\newcommand{\araa}{ARAA}

\newcommand{\apss}{ApSS}

\newcommand{\nat}{Nature}

\newcommand{\physscr}{Phys.\ Scr.}

\title[The magnetic field of HD\,144941]{Detection of an extremely strong magnetic field in the double-degenerate binary merger product HD\,144941}
\author[M.\ E.\ Shultz]
{M.\ E.\ Shultz$^{1}$\thanks{E-mail: mshultz@udel.edu},
O.\ Kochukhov$^2$, 
J.\ Labadie-Bartz$^3$, 
A.\ David-Uraz$^{4,5}$,
and S.\ P.\ Owocki$^{1}$\\
$^1$Department of Physics and Astronomy, University of Delaware, 217 Sharp Lab, Newark, Delaware, 19716, USA\\
$^2$Department of Physics and Astronomy, Uppsala University, Box 516, Uppsala 75120, Sweden\\
$^3$Instituto de Astronomia, Geofísica e Ci\^encias Atmosf\'ericas, Universidade de S\~ao Paulo, Rua do Mat\~ao 1226, Cidade Universit\'aria,\\
05508-900 S\~ao Paulo, SP, Brazil\\
$^4$Department of Physics and Astronomy, Howard University, Washington, DC 20059, USA\\
$^5$Center for Research and Exploration in Space Science and Technology, and X-ray Astrophysics Laboratory, NASA/GSFC, Greenbelt,\\
MD 20771, USA\\
}

\begin{document}

\date{}

\pagerange{\pageref{firstpage}--\pageref{lastpage}} \pubyear{2018}

\maketitle

\label{firstpage}

\begin{abstract}
HD\,144941 is an extreme He (EHe) star, a rare class of subdwarf OB star formed from the merger of two white dwarf (WD) stars. Uniquely amongst EHe stars, its light curve has been reported to be modulated entirely by rotation, suggesting the presence of a magnetic field. Here we report the first high-resolution spectropolarimetric observations of HD\,144941, in which we detect an extremely strong magnetic field both in circular polarization (with a line-of-sight magnetic field averaged over the stellar disk \bz~$\sim -8$~kG) and in Zeeman splitting of spectral lines (yielding a magnetic modulus of \bmodulus~$\sim 17$~kG). We also report for the first time weak H$\alpha$ emission consistent with an origin an a Centrifugal Magnetosphere (CM). HD\,144941's atmospheric parameters could be consistent with either a subdwarf or a main sequence (MS) star, and its surface abundances are neither similar to other EHe stars nor to He-strong magnetic stars. However, its H$\alpha$ emission properties can only be reproduced if its mass is around 1 \msun, indicating that it must be a post-MS object. Since there is no indication of binarity, it is unlikely to be a stripped star, and was therefore most likely produced in a WD merger. HD\,144941 is therefore further evidence that mergers are a viable pathway for the generation of fossil magnetic fields. 
\end{abstract}


\section{Introduction}

Magnetic fields exist on about 10\% of hot main sequence (MS) OBA stars \citep{2019MNRAS.483.2300S,2017MNRAS.465.2432G}. The magnetic fields of hot stars are generally strong \citep[a few hundred G to about 30 kG;][]{2019MNRAS.490..274S}, globally organized and topologically simple \citep[typically dipolar;][]{2018MNRAS.475.5144S}, and stable over at least thousands of rotational cycles \citep{2018MNRAS.475.5144S}. No dynamo mechanism is known to be able to generate strong, globally organized magnetic fields within a radiative envelope, and there is furthermore no correlation between magnetic field strength and rotation \citep{2019MNRAS.490..274S}, as is observed for the dynamo-generated magnetic fields of stars with convective envelopes \citep{2016MNRAS.457..580F}. For these reasons the magnetic fields of hot stars are thought to be fossil magnetic fields: leftovers from a previous era in the star's evolution. 

While the stability of fossil fields within radiative envelopes over evolutionary timescales has been demonstrated by magnetohydrodynamic (MHD) simulations \citep{2004Natur.431..819B}, their origin remains a subject of debate. Broadly, there are two competing mechanisms. The first involves amplification of a seed magnetic field from the star-forming material by a pre-MS convective dynamo \citep{2015IAUS..305...61N}. The second involves dynamos powered by binary mergers \citep{2009MNRAS.400L..71F}. There are several lines of evidence supporting the merger scenario. First, the expected merger rate for massive stars is similar to the 10\% incidence of fossil fields on the upper MS \citep{2012Sci...337..444S}. Second, the majority of magnetic Chemically Peculiar (mCP) stars are single: in contrast to the high binary fraction amongst the general population of hot stars \citep{2012Sci...337..444S}, only about 2\% of hot close binaries (with orbital periods less than 30 d) contain a magnetic star \citep{2015IAUS..307..330A}. Finally, MHD simulations have demonstrated the plausibility of this mechanism \citep{2019Natur.574..211S}. However, no magnetic field has ever been detected in an early-type star that is unambiguously a merger product. 

HD\,144941 is an Extreme Helium (EHe) star \citep{1997A&A...323..177H}, a rare variety of hot subdwarf with He surface abundances in excess of 90\%. EHe stars are short-lived, rapidly evolving products of mergers of two white dwarf (WD) stars \citep{2002MNRAS.333..121S}, a scenario established based on the fact that they are never found in close binaries \citep{2002MNRAS.333..121S}; the excellent agreement between their surface abundances and the abundance patterns expected from WD mergers \citep{2002MNRAS.333..121S}; and their rarity, which matches the expected fraction given the WD merger rate and the expected lifetime of a merger product before it evolves back down onto the WD cooling track \citep{2002MNRAS.333..121S,2001A&A...365..491N}. 

Analysis of HD\,144941's {\em Kepler-2} ({\em K2}) light curve revealed perfectly periodic modulation with a period of about 14 days \citep{2018MNRAS.475L.122J}. This is much too long to be ascribed to the pulsations that are often found in EHe stars \citep{2020MNRAS.495L.135J}, and indeed HD\,144941 shows no sign of pulsational variation on short timescales \citep{2018MNRAS.475L.122J}. The light curve is also very complex, with numerous harmonics in the periodogram. Its shape is inconsistent with binary signals such as eclipses. The photometric modulation was therefore interpreted as rotational modulation, similar to that almost invariably seen in mCP stars. In the case of mCP stars, atmospheric stabilization by strong magnetic fields enables the accumulation of long-lived surface chemical abundance patches via atomic diffusion \citep{2019MNRAS.482.4519A}, which produce perfectly periodic photometric variability associated with the star's rotation \citep{2012A&A...537A..14K}. Since no other mechanism is known to be capable of stabilizing the radiative envelope of a hot star, this suggested that HD\,144941 may host a magnetic field. 

To test this hypothesis, we acquired 2 circularly polarized (Stokes $V$) spectra with the high-resolution ESPaDOnS (Echelle SpectroPolarimetric Device for the Observation of Stars) spectropolarimeter \citep{d1997} at the 3.6-m Canada-France-Hawaii Telescope (CFHT). The ESPaDOnS observations, together with ancillary spectroscopic and photometric data utilized in the analysis, are described in \S~\ref{sec:observations}. In \S~\ref{sec:rot} the rotation period is revisited. The magnetic analysis is described in \S~\ref{sec:mag}. Spectroscopic line profile variability is examined in \S~\ref{sec:lpv}. In \S~\ref{sec:ehe_star} the question of the whether HD\,144941 is an EHe star or a misidentified He-strong (He-s) mCP star is examined via analysis of the star's atmospheric and fundamental parameters, surface abundances, and H$\alpha$ emission properties. The results are discussed in \S~\ref{sec:discussion} and conclusions summarized in \S~\ref{sec:conclusions}.

\section{Observations}\label{sec:observations}

\begin{figure*}
\includegraphics[width=\textwidth]{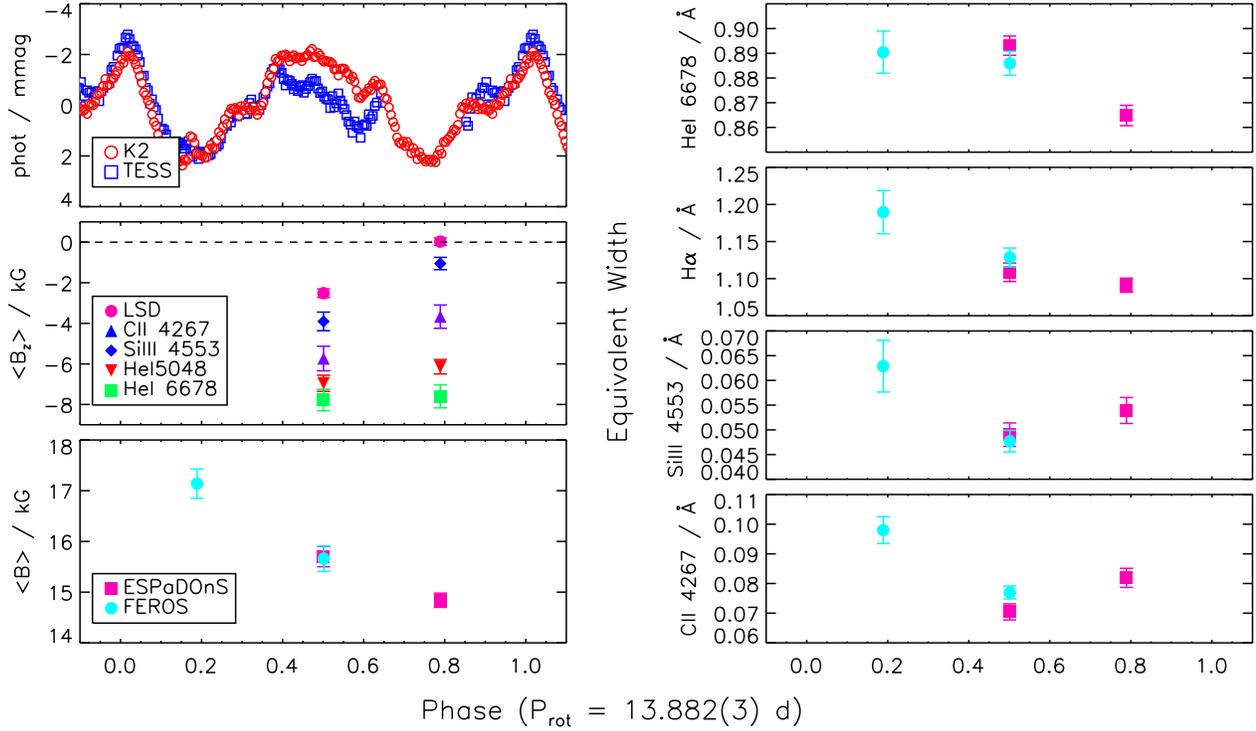}
\caption{Rotational diagnostics of HD 144941 phased with the rotation period. {\em Left panels}: {\em top}: Rotational-phase-binned {\em K2} and {\em TESS} light curves; {\em middle}: ESPaDOnS \bz~measurements obtained from various spectral lines and from LSD profiles; {\em bottom}: magnetic modulus $\langle B \rangle$ measured from ESPaDONS and FEROS spectra. Note that the two light curves agree well everywhere except between phases 0.4 and 0.6. Note also the considerable variation in \bz~as determined from different chemical elements. {\em Right panels}: EW measurements of 4 different spectral lines from ESPaDOnS and FEROS spectra. Variability is present in all lines, about 2\% in He, 17\% in H, and around 35\% in both Si and C. The ESPaDOnS and FEROS measurements obtained near phase 0.5 agree well in all cases.}
\label{hd144941_halpha_bz_phot}
\end{figure*}

\begin{figure*}
\includegraphics[width=\textwidth]{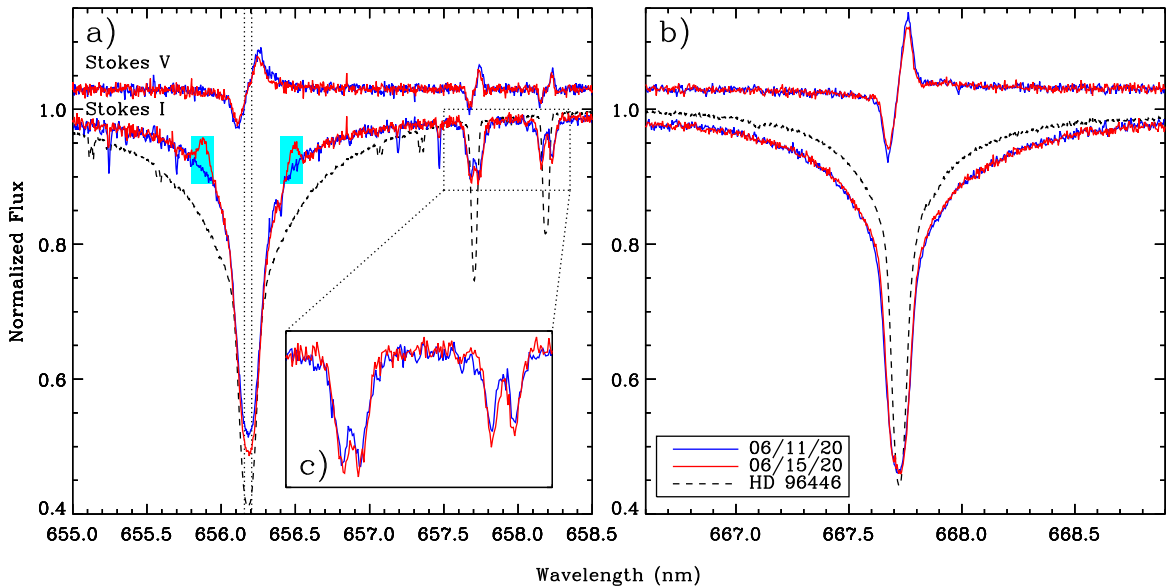}
\caption{ESPaDOnS observations of HD 144941 in the vicinity of (a) the H$\alpha$ line, (b) the He~{\sc I}~667.8~nm line, and (c) the C~{\sc II} doublet. Stokes $V$ is shown without amplification, with the Stokes $V$ continuum shifted for display purposes. For comparison, the spectrum of the He-s Bp star HD 96446 is shown. Note HD 144941's much stronger He, and much weaker H, lines. In (a) magnetospheric H$\alpha$ emission is highlighted with shaded cyan regions. Vertical dotted lines indicate $\pm v\sin{i}$. Zeeman splitting is clearly detectable in both C~{\sc II} lines (c). }
	\label{specplot}
\end{figure*}

\begin{table}
\centering
\caption[]{Observation log and magnetic measurements. Inst. refers to the instrument, F(EROS) or E(SPaDOnS). \bz~measurements are from the He~{\sc i} 667.8 nm line, \bmodulus~from the C~{\sc ii} doublet.}
\label{obslog}
\begin{tabular}{l l c r r}
\hline\hline
Date & HJD & Inst. & \bz & \bmodulus \\
     & $-2450000$ & & (kG) & (kG) \\
\hline
Feb 25 2005 & 3426.9 & F & -- & $17.14 \pm 0.29$ \\
Apr 08 2006 & 3833.8 & F  & -- & $15.66 \pm 0.25$ \\
Jun 11 2020 & 9011.8 & E & $-7.78 \pm 0.53$ & $15.70 \pm 0.20$ \\
Jun 15 2020 & 9015.8 & E & $-7.60 \pm 0.56$ & $14.84 \pm 0.14$ \\
\hline\hline
\end{tabular}
\end{table}

\subsection{Space photometry} 

The previously analyzed \citep{2018MNRAS.475L.122J} \citep{2014PASP..126..398H} ({\em K2}) light curve was obtained from the Mikulski Archive for Space Telescopes (MAST). The NASA {\em Kepler} satellite is a $\mu$mag-precision space photometer with a 110 square degree field of view operating in the 400 to 850 nm bandpass, intended for high-cadence, long-duration observations with the goal of detecting transiting exoplanets. The {\em K2} mission was an extension of the original {\em Kepler} mission, following the failure of two of the satellite's reaction wheels; by utilizing pressure from the solar wind, the satellite could be stabilized on a given field of view for about 3 months, enabling it to observe fields along the ecliptic. HD\,144941 was observed in long-cadence mode (1 observation every 30 min) between Aug and Nov of 2014 \citep{2018MNRAS.475L.122J}.

HD\,144941 was observed by the Transiting Exoplanet Survery Satellite ({\em TESS}) \citep{2015JATIS...1a4003R} in sector 12 between May 21, 2019 and June 18, 2019. The data were acquired from the MAST archive. {\em TESS} uses four cameras with a total field of view of $24^\circ \times 96^\circ$, with a bandpass covering 600 to 1050 nm. The initial two-year {\em TESS} mission began in 2018, during which it completed coverage of almost the entire sky. During each year, 13 sectors were observed for 27 days each, with a nominal precision of 60 ppm hr$^{-1}$ (although this varies between fields and targets). High-priority targets are observed with a two-minute cadence, and the processed light curves made available on MAST immediately following reduction; HD\,144941 falls into this category.

\subsection{ESPaDOnS spectropolarimetry} 

The Echelle SPectropolarimetric Device for the Observation of Stars (ESPaDOnS) spectropolarimeter \citep{d1997} is a high-resolution ($R \sim 65,000$) \'echelle spectropolarimeter mounted at the Cassegrain focus of the Canada-France-Hawaii Telescope. It covers the spectrum between about 370 nm and 1000 nm across 40 overlapping spectral orders. Each observation consists of 4 sequential sub-exposures obtained with different polarisations, which are combined to create one circular polarisation (Stokes $V$) and 2 diagnostic null ($N$) spectra in which intrinsic polarisation from the source is cancelled out \citep{d1997,2009PASP..121..993B}. Stokes $V$ is sensitive to the line-of-sight magnetic field, while the $N$ spectra are used to characterize photon noise and check for nominal instrument performance. The data were reduced wth the standard Upena pipeline \citep{2011tfa..confE..63M}, which is based on the Libre-ESpRIT pipeline \citep{d1997}. Instrument operation, performance, and data reduction and characteristics are described in detail by \cite{2016MNRAS.456....2W}.

HD\,144941 was observed on two nights with ESPaDOnS, June 11 and June 15, 2020. On each night two spectropolarimetric sequences (i.e. 8 sub-exposures) were taken, and the two sequences co-added to increase the signal-to-noise ($S/N$). Each sub-exposure was 867 s long (or 1.93 hr for the full spectropolarimetric sequence), and the final co-added spectra had peak per-pixel $S/N$ of 303 and 315, respectively. 

\subsection{FEROS spectroscopy} 

The Fiberfed Extended Range Optical Spectrograph (FEROS) \citep{1999Msngr..95....8K} is a high-resolution ($R \sim 48,000$) \'echelle spectrograph mounted at the La Silla Observatory 2.2 m telescope. It covers the spectrum from 350 to 920 nm across 39 overlapping orders. Two observations of HD\,144941 were acquired, on Feb 25, 2005 and Apr 8, 2006. The reduced data were downloaded from the European Southern Observatory archive. 

\section{Rotation Period}\label{sec:rot}

The published rotational period of $13.9 \pm 0.2$~d \citep{2018MNRAS.475L.122J} does not satisfactorally phase the {\em K2} and {\em TESS} data, indicating that the period is not sufficiently precise to phase data outside of the {\em K2} window. 

As a first attempt to improve the period, the {\em K2} and {\em TESS} datasets were combined into a single dataset and analyzed using the standard period analysis program PERIOD04 \citep{2005CoAst.146...53L}. This did not provide useful results, due to the incomplete sampling of the {\em TESS} light curve (there is a 3-day gap in the middle of dataset), the obvious differences between the {\em TESS} and {\em K2} light curves (see Fig. \ref{hd144941_halpha_bz_phot}), and the greater scatter in the {\em TESS} data. The differences in the phase-binned light-curves are likely due to the differences between the {\em Kepler} and {\em TESS} bandpasses. The shapes of mCP light curves frequently differ between bandbasses \citep{2012A&A...537A..14K,2015A&A...576A..82K,2015A&A...584A..17P}.

As a next step, we analyzed the {\em K-2} light curve independently, optimizing the period by simultaneously fitting the rotational frequency $f_{\rm rot}$ and its 8 significant harmonics \citep{2018MNRAS.475L.122J}, which were fixed to integer multiples of $f_{\rm rot}$. This yielded a period of $13.89(2)$~d, consistent with the period given by \cite{2018MNRAS.475L.122J} but somewhat more precise \citep[the number in brackets gives the analytic uncertainty in the least significant digit;][]{1976fats.book.....B}. We then conducted a Lomb-Scargle analysis of the combined dataset, centering the frequency window on $f_{\rm rot}$ and restricting the width to three times the formal uncertainty in order to avoid the severe aliasing resulting from the considerable time gap between the {\em K2} and {\em TESS} datasets. As before, the rotational frequency was then optimized by fitting $f_{\rm rot}$ and its 8 harmonics. This gave $P_{\rm rot} = 13.882(3)$~d. The phase-binned light curves are shown folded with this period in Fig.\ \ref{hd144941_halpha_bz_phot}.


The zero-point for the ephemeris $T0 = {\rm HJD}~2456880.9(1)$ is the heliocentric julian date of maximum light one rotational cycle before the first {\em K2} observation.

\section{Magnetic field}\label{sec:mag}

The ESPaDOnS observations are shown in Fig.\ \ref{specplot} in the vicinity of the H$\alpha$ line and the He~{\sc i}~667.8~nm line. In both observations, a Zeeman signature is clearly detected in Stokes $V$. HD 144941 is therefore without doubt a magnetic star. 

\subsection{Least-Squares Deconvolution} 

\begin{figure}
\includegraphics[width=0.45\textwidth]{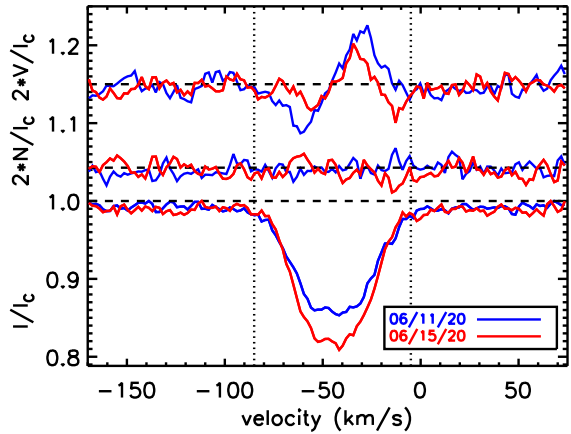}
\caption{LSD Stokes $I$ (bottom), $N$ (middle), and Stokes $V$ (top) profiles. Note that the $N$ and Stokes $V$ profiles have been amplified by a factor of 2 and offset vertically. Vertical dotted lines show the integration range for measurement of \bz. Stokes $I$ profiles show a change in line strength, likely due to chemical spots. In contrast to the results for He and H lines (Fig. \ref{specplot}) Stokes $V$ profiles show a large change in morphology. }
\label{lsd}
\end{figure}

Least-squares deconvolution (LSD) profiles are mean line profiles obtained via combining a line mask, consisting of line intensities and Land\'e factors, with the observed spectrum, and deconvolving the mean line profile \citep{d1997}. The iLSD package was used for this operation \citep{koch2010}. LSD profiles were extracted using scaling factors of 1.2 in Land\'e factor, 0.1 in line depth, and 500 nm for wavelength \citep{koch2010}. 

The line list was obtained from the Vienna Atomic Line Database (VALD3) \citep{piskunov1995, ryabchikova1997, kupka1999, kupka2000,2015PhyS...90e4005R} using an `extract stellar' request for a star with solar metallicity and atmospheric parameters corresponding to those of HD\,144941 \citep{2017ApJ...847..127P}. Solar metallicity was used rather than the star's much lower, measured abundances so as to ensure all possible lines are available for analysis. Following the usual procedure \citep{2018MNRAS.475.5144S}, the line mask was cleaned to remove lines blended with strong H or He lines; lines obscured by telluric features; interstellar lines; lines in parts of the spectrum affected by rippling; and lines absent from the observed spectrum. This ensures that the mean Stokes $I$ and $V$ profiles are obtained only from lines formed in the stellar atmosphere and possessing similar line profiles. The depths of the remaining 50 lines were then manually adjusted to match the line depths of the observed spectrum. The strongest contributions are from Si~{\sc iii} lines. The resulting LSD profiles are shown in Fig.\ \ref{lsd}.

\subsection{Longitudinal Magnetic Field} 

The longitudinal surface magnetic field averaged over the stellar disk \bz \citep{mat1989} was measured first from a selection of individual lines, with rest wavelengths and Land\'e factors obtained from the VALD line list. All lines were chosen as being relatively strong and isolated. Results from 4 representative lines are shown in Fig.\ \ref{hd144941_halpha_bz_phot}: C~{\sc II}~426.7, Si~{\sc III}~455.3, He~{\sc I}~504.8, and He~{\sc I}~667.8~nm. In addition, \bz~was measured using Si~{\sc III}~456.8 and He~{\sc I}~443.8~nm. Integration ranges were chosen based on the widths of individual lines. For measurement of \bz~from LSD profiles, a Land\'e factor of 1.2 was used, corresponding to the value used for extracting LSD profiles.

While a magnetic field is clearly detected in both observations, \bz~is discrepant between lines and especially between elements. In all cases, the strongest \bz~was measured on June 11$^{\rm th}$, ranging from $-7.8 \pm 0.5$~kG from He~{\sc I}~667.8~nm, to $-3.9 \pm 0.5$~kG from Si~{\sc III}~455.3~nm. The second observation on June 15$^{\rm th}$ yielded no significant change when measured using He lines ($-7.6 \pm 0.5$~kG using He~{\sc I}~667.8~nm), but significant changes in metallic lines, with the largest difference in Si~{\sc III}~455.3~nm, to $-1.0 \pm 0.3$~kG. The results from Si~{\sc III}~456.8~nm are consistent with those from Si~{\sc III}~455.3~nm, while the results from He~{\sc I}~443.8~nm are consistent with the other two He~{\sc I} lines. The C~{\sc II}~426.7~nm line yielded intermediate results; note that this line is an unresolved triplet, with Land\'e factors of 0.9, 1.029, and 1.071, for which the mean value was used. Measurement of \bz~from the C~{\sc ii} doublet at 657.8 and 658.3~nm yields results similar to those of other metallic lines, i.e.\ the strongest \bz~on June 11$^{\rm th}$ and a significantly weaker \bz~on June 15$^{\rm th}$.

\bz~measured from the LSD profiles yields $-2.5 \pm 0.2$~kG from the first observation and $0.0 \pm 0.2$ kG from the second (see Fig.\ \ref{hd144941_halpha_bz_phot}). Examination of the LSD Stokes V profiles, shown in Fig.\ \ref{lsd}, reveals the reason for this discrepancy with the results from individual He lines: whereas the He Stokes $V$ profiles of the two observations are little changed from one another, with the exception that the second is slightly weaker than the first (Fig. \ref{specplot}), the LSD Stokes $V$ profiles show large changes. In particular the second observation is a crossover signature: hence, \bz~is consistent with 0, but the magnetic field is nevertheless clearly detected \citep{1995A&A...293..733M}. 

Considerable differences in \bz~as measured from different elements are often seen in mCP stars \citep{2018MNRAS.475.5144S}, and arise due to the interaction of chemical spots with the surface magnetic field \citep{2015MNRAS.447.1418Y}. Inhomogenous surface chemical abundance distributions lead to differential weighting of both unpolarised and polarised flux from the stellar atmosphere, thereby warping magnetic diagnostics. Since different chemical elements are not typically distributed in the same way across the surface of a given star \citep{2014MNRAS.444.1442S,2015MNRAS.447.1418Y,2015A&A...574A..79K,2016A&A...588A.138R,2017A&A...605A..13K,2018A&A...609A..88R}, magnetic diagnostics therefore differ between elements, and a secure model of the surface magnetic field can only be obtained with simultaneous reconstruction of the chemical abundance distribution and magnetic field with tomographic mapping techniques \citep{2014MNRAS.444.1442S,2015MNRAS.447.1418Y,2015A&A...574A..79K,2016A&A...588A.138R,2017A&A...605A..13K,2018A&A...609A..88R}. 

There is some indication that stars with magnetic fields departing significantly from a purely dipolar geometry exhibit larger discrepancies in \bz~between results from different elements than do stars with more dipolar fields. One extreme case is Landstreet's Star (HD 37776), which exhibits the largest differences between single-element measurements amongst the population of magnetic early B-type stars \citep{2018MNRAS.475.5144S}, and possesses one of the most complex surface magnetic field topologies known amongst mCP stars \citep{koch2010,2019A&A...621A..47K}. The large discrepancies in single-element \bz~measurements for HD\,144941 may be an indication that the star's surface magnetic field is not well described by a dipole, in contrast to the majority of fossil magnetic fields \citep{2019A&A...621A..47K}, although a much larger spectropolarimetric dataset is required to constrain this hypothesis.  


\subsection{Mean Magnetic Field Modulus} 

\begin{figure}
\includegraphics[width=0.45\textwidth]{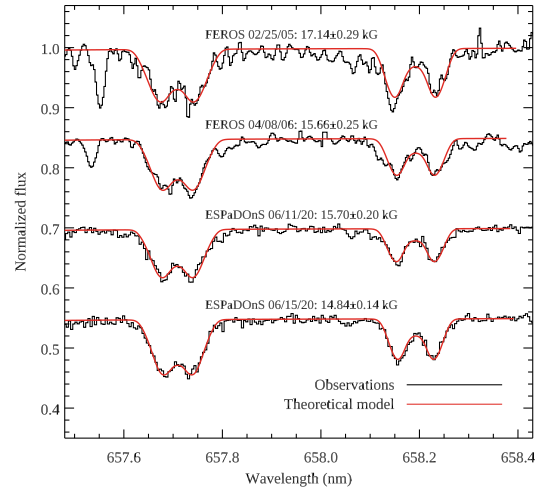}
\caption{Comparison of the observed high-resolution spectra of HD 144941 (histogram) around the C II $\lambda$ 657.805 and 658.288 nm lines and theoretical spectrum synthesis calculations (red solid lines). Observations and model spectra corresponding to different dates are shifted vertically. The best fitting field modulus is indicated next to each pair of spectra.}
\label{cii_bmod}
\end{figure}

As can be seen in the inset in Fig.\ \ref{specplot}, Zeeman splitting is clearly detected in the C~{\sc ii} doublet in the red wing of the H$\alpha$ line. A complementary magnetic field diagnostic, the mean magnetic field modulus \bmodulus~was therefore determined from the C {\sc ii}~657.805 and 658.288~nm lines. This provides a measure of the average surface magnetic field. The C~{\sc ii} lines were chosen for modelling because they represent the clearest instance of resolved Zeeman-split lines in the available high-resolution spectra of HD\,144941. Determination of $\langle B \rangle$ involved calculation of theoretical spectra with a polarised radiative transfer code \citep{koch2010} using model atmosphere parameters and abundances \citep{2017ApJ...847..127P} appropriate for HD 144941. The line list, including Land\'e factors necessary for calculation of Zeeman splitting of the C {\sc II} lines, was obtained from the VALD database \citep{piskunov1995, ryabchikova1997, kupka1999, kupka2000,2015PhyS...90e4005R}. The stellar surface magnetic field was approximated by a uniform radial magnetic structure. Each of the four high-resolution spectra were fitted individually, letting a least-squares algorithm adjust the field strength, the C abundance, the projected rotational velocity, and the stellar radial velocity. The resulting best-fitting spectra are illustrated in Fig.\ \ref{cii_bmod}. For the two ESPaDOnS observations we derived $\langle B \rangle = 15.70 \pm 0.20$ and $14.84 \pm 0.14$~kG. Analysis of the two FEROS spectra yielded $\langle B \rangle = 17.14  \pm 0.29$ and $15.66 \pm 0.25$~kG. An average projected rotational velocity of $11.2 \pm 0.7$ km/s, along with the radial velocity $-43.7 \pm 0.3$~km/s, were deduced. No evidence of changes in the radial velocity was found, indicating that HD\,144941 is not a member of a binary system. 

 $\langle B \rangle$ is shown phased with the rotation period in Fig.\ \ref{hd144941_halpha_bz_phot}; note that a FEROS and ESPaDOnS observation were obtained at a similar phase, about 0.5, and agree well with one another, providing an independent test of the rotation period.

\section{Line Profile Variability}\label{sec:lpv}

Photometric rotational variation in hot stars is usually attributed to surface chemical abundance patches. Chemical spots have been shown to be capable, in principle, of leading to rotational modulation in subdwarf stars as well \citep{2021arXiv210404117K}. In that case we also expect that chemical spots will lead to spectroscopic line profile variations. As a quantitative measure of spectroscopic variation, equivalent widths (EW) were measured from ESPaDonS and FEROS spectra by first renormalizing to continuum regions on either side of the line profile using linear fits, and then calculating the integrated flux inside the line profile. 

The bottom panels of Fig.\ \ref{hd144941_halpha_bz_phot} show EW measurements of 4 selected lines from ESPaDOnS and FEROS spectra, folded with the rotation period determined above. One of the FEROS measurements was obtained at a similar rotational phase as one of the ESPaDOnS measurements, as determined with the updated ephemeris. In all lines, there is a good agreement in EW between these two ESPaDOnS and FEROS observations, serving as an independent verification of the photometrically determined rotational period. 

While all lines are variable, there is a considerable difference in the relative amplitude (i.e., the ratio of the amplitude to the mean EW) of variation between different chemical species. The least variable line is He~{\sc I}~667.8~nm, which varies at about the 2\% level. C~{\sc II}~526.7~nm and Si~{\sc III}~455.3~nm, however, have relative amplitudes around 36\%. H$\alpha$ is intermediate, around 17\%; however as explored below, the variability in H$\alpha$ is likely to originate in the magnetosphere rather than in photospheric chemical spots as with He and metallic lines. Notably, the star's light curve is highly structured, showing a large number of harmonics of the rotational frequency \citep{2018MNRAS.475L.122J}. While surface spots are able to reproduce the fundamental and lowest harmonics, the higher harmonics are difficult to reproduce with photospheric features, leading to the suggestion that they originate in the circumstellar environment \citep{2021arXiv210404117K}, a prediction which is consistent with the detection of H$\alpha$ emission.

The very low amplitude of variation seen in HD 144941's He lines is highly anomalous as compared to the He lines of He-s mCP stars, which are in general highly variable, ranging on the low end from relative amplitudes of around 16\% for HR 5907 \citep{grun2012} to up to around 80\% for HR 7355 \citep{rivi2013}. The well-known He-s stars $\sigma$ Ori E and HD 184927 both have relative amplitudes of around 30\% \citep{oks2012, 2015MNRAS.447.1418Y}. The one exception to this is HD\,96446, which has the highest surface He number fraction of the He-s stars(60\%) and shows an amplitude of variation of just 3\% in He~{\sc i} lines \citep{2019A&A...626A..94G}. This may indicate that, in contrast to the majority of He-s mCP stars, the distribution of He in HD 144941's photosphere is nearly homogeneous. On the other hand, this may be a trivial consequence of He being the dominant element: H EW variation is similarly low in most He-s stars, since in those cases H is the dominant element. The similarly low He EW variation in HD\,96446 points to this interpretation.

\section{Is HD 144941 an EHe Star or a Misidentified He-strong mCP Star?}\label{sec:ehe_star}

The conclusion that HD 144941 is in fact a merger product rests upon its identification as an EHe star. The chemical peculiarity of HD 144941 as compared to other EHe stars has been much remarked upon \citep{2017ApJ...847..127P}. The possibility that it might be a He-s mCP star was previously rejected on the basis of its {\em Gaia} Data Release 1 (DR1) parallax \citep{2018MNRAS.475L.122J}, which yielded a luminosity below the MS. However, its {\em Gaia} Early DR3 \citep{2020arXiv201202061G} parallax, $\pi = 0.67 \pm 0.03$~mas, yields a higher luminosity, which could be consistent with a MS star, reopening the question of whether HD 144941 has been misidentified as a post-MS object. In the following, we examine HD 144941's luminosity, surface gravity, and surface chemical abundances, comparing them to the properties of both EHe stars and mCP stars, in order to determine to which class the star is most likely to belong. In the subsequent section, the star's magnetospheric H$\alpha$ emission is also utilized as an independent constraint on the mass. Effective temperatures \teff~and surface gravities $\log{g}$ for all EHe stars were obtained from published values based on spectroscopic modelling \citep{2006ApJ...638..454P,2011ApJ...727..122P,2017ApJ...847..127P}. 

\subsection{Stellar parameters}\label{stellarpar}

\begin{figure}
\includegraphics[width=0.5\textwidth]{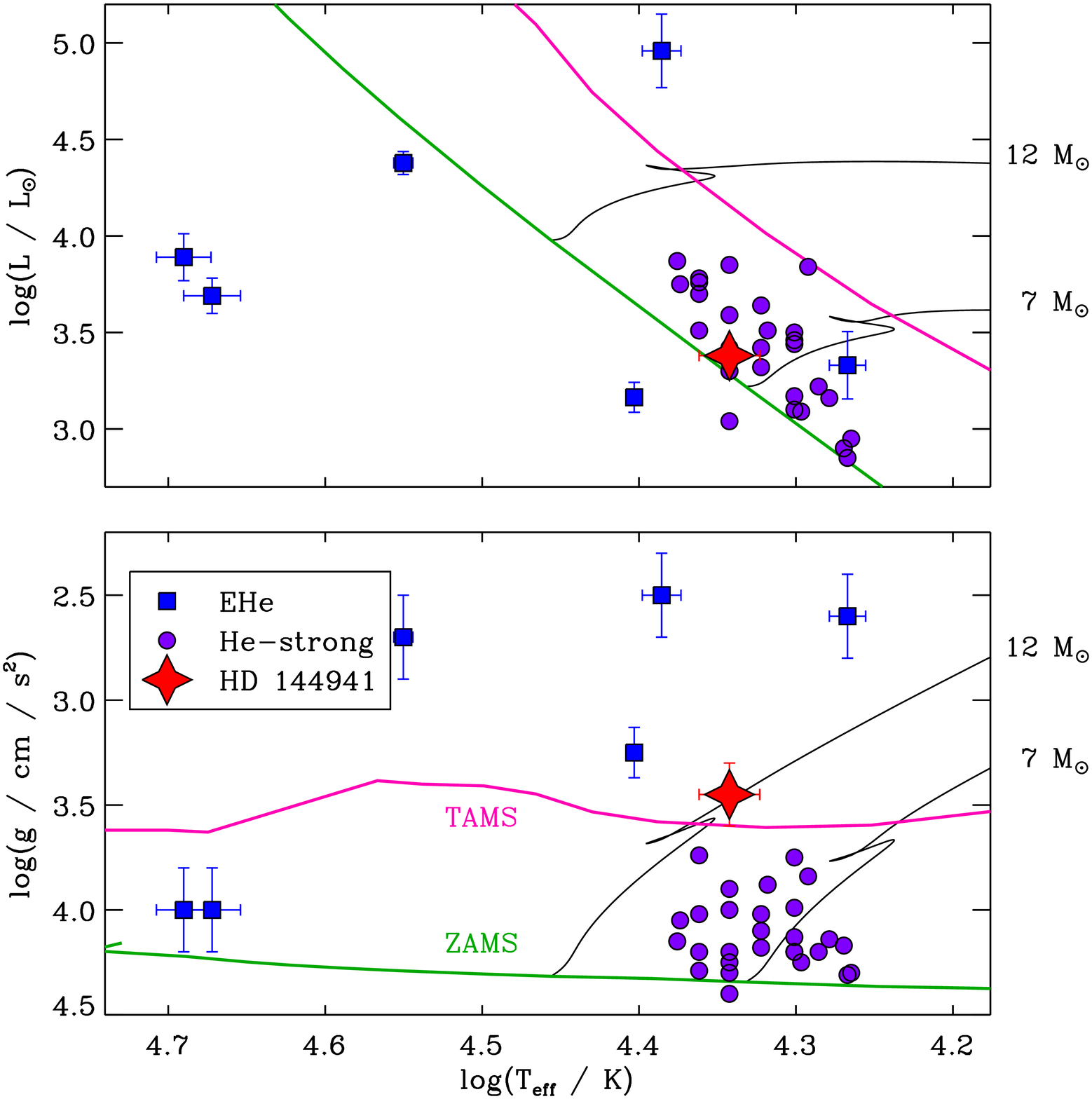}
\caption{Hertzsprung-Russell diagram ({\em top}) and Kiel diagram ({\em bottom}) showing the positions of EHe stars, He-s stars, and HD\,144941. While HD\,144941's position on the HRD could be consistent with a MS He-s star, the surface gravities of EHe stars, HD\,144941 included, are systematically lower than implied by their luminosities, as compared to MS models.}
\label{hrd}
\end{figure}

The {\em Gaia} Early Data Release 3 (EDR3) parallax is $\pi = 0.67 \pm 0.03$~mas, giving a distance of $1490 \pm 70$~pc. The literature gives reddening $E(B-V)$ between 0.25 and 0.31 \citep{1986ESASP.263..369J, 2010MNRAS.404.1698J}. However, the highest value of $E(B-V) = 0.31$ was obtained using \teff~$= 27$~kK \citep{2010MNRAS.404.1698J}, much higher than is consistent with the value obtained from NLTE modelling of visible band spectroscopy, about 22 kK \citep{2017ApJ...847..127P}. Comparison of dereddened colours computed with the star's UBVJHK magnitudes to empirical calibrations \citep{2013ApJS..208....9P} yields $E(B-V) = 0.21$. Reddening can also be determined using the Stilism tomographic dust map \citep{2014A&A...561A..91L, 2017A&A...606A..65C}. While the Stilism sightline in the direction of HD 144941 extends only to 950 pc, the increase in $E(B-V)$ is linear from 300 pc out to this distance, and extrapolation of this trend to 1490 pc yields exactly the value obtained from the photometric colours. We therefore adopted $E(B-V) = 0.21$, with extinction $A_{\rm V} = R_{\rm V} E(B-V) = 0.65$ using the usual reddening law $R_{\rm V} = 3.1$.

For the bolometric correction $BC$ we used two calibrations, one for MS OB stars \citep{nieva2013} and another for mCP stars \citep{2008AA...491..545N}, which for a 22 kK star yield $BC$ between $-2.32$ and $-2.12$ mag. With the star's distance, extinction, and $V$ magnitude, the luminosity is then $\log(L_*/{\rm L_\odot}) = 3.38 \pm 0.07$, and its radius is $R_* = 3.4 \pm 0.3$~\rsun. Placing the star on the Hertzsprung-Russell Diagram (HRD), and comparing to Geneva evolutionary tracks for MS stars \citep{ekstrom2012}, if the star is on the MS it should have a mass of about 8~\msun.

In Fig.\ \ref{hrd} HD\,144941's position on the HRD is compared to that of MS He-s stars \citep{2019MNRAS.485.1508S} and to that of other EHe stars \citep{2006ApJ...638..454P,2011ApJ...727..122P,2017ApJ...847..127P}, where luminosities were determined from {\em Gaia} EDR3 parallaxes in the same fashion as for HD\,144941. With the exception of HD\,144941 and one other EHe star, the majority of the EHe stars are inconsistent with the MS, being either under- or overluminous.

\subsection{Surface gravity and mass}\label{specmass}

Surface gravity measurements vary widely: $\log{g} = 3.45 \pm 0.15$ \citep{2017ApJ...847..127P}, $3.9 \pm 0.2$ \citep{1997A&A...323..177H}, and $4.15 \pm 0.1$ \citep{2005A&A...443L..25P}. Using the radius determined with the {\em Gaia} EDR3 parallax, the spectroscopic mass is then $1.2 \pm 0.5$~\msun~using the lowest $\log{g}$, and $6 \pm 2$~\msun~using the highest value. The highest value is formally consistent with the mass of a MS star at HD 144941's position on the HRD, but the lowest value is clearly inconstent with a MS star. We note that the lowest value is from the more recent and most careful analysis to date, obtained using both non-Local Thermodynamic Equilibrium (NLTE) model atmospheres and model atoms \citep{2017ApJ...847..127P}, while the highest value was obtained using a hybrid method of LTE model atmospheres and NLTE model atoms \citep{2005A&A...443L..25P}.

The lower panel of Fig.\ \ref{hrd} shows HD\,144941's position on the Kiel diagram using the NLTE $\log{g}$ \citep{2017ApJ...847..127P}, as compared to that of MS He-s stars \citep{2019MNRAS.485.1508S} and EHe stars \citep{2011ApJ...727..122P,2017ApJ...847..127P}. While the surface gravities of the He-s stars are consistent with their luminosities, EHe stars have systematically lower surface gravities than would be inferred from their luminosities. HD\,144941 is no exception to this general trend. However, comparing HD\,144941 to the two EHe stars with the most comparable luminosities and effective temperatures, HD\,144941 has the highest surface gravity. This indicates that HD\,144941 must be more massive than most EHe stars. Indeed, the spectroscopic masses of the 6 other EHe stars in Fig.\ \ref{hrd} range from 0.25 to 0.6 M$_\odot$.

The surface gravity of hot stars is inferred primarily from the pressure broadening of H Balmer lines and He lines. Strong magnetic fields can contribute magnetic pressure via the Lorentz force, with the result that the flux in H and He lines can be changed by up to several percent locally and around 1.5\% when averaged over the stellar disk \citep{2004A&A...420..993V,2010A&A...509A..28S,2014Ap&SS.352...95V}, thus modifying the inferred surface gravity by up to 0.1 dex. This is equivalent to about a 20\% systematic uncertainty in the spectroscopic mass. Therefore, while the highest published value of $\log{g}$ could be consistent with a mass as high as 9.6~\msun, the lowest published value could be consistent with a mass as low as 0.56~\msun, comparable to the upper range observed in EHe stars.

\subsection{Chemical Abundances}\label{abund}

\begin{figure*}
\includegraphics[width=\textwidth]{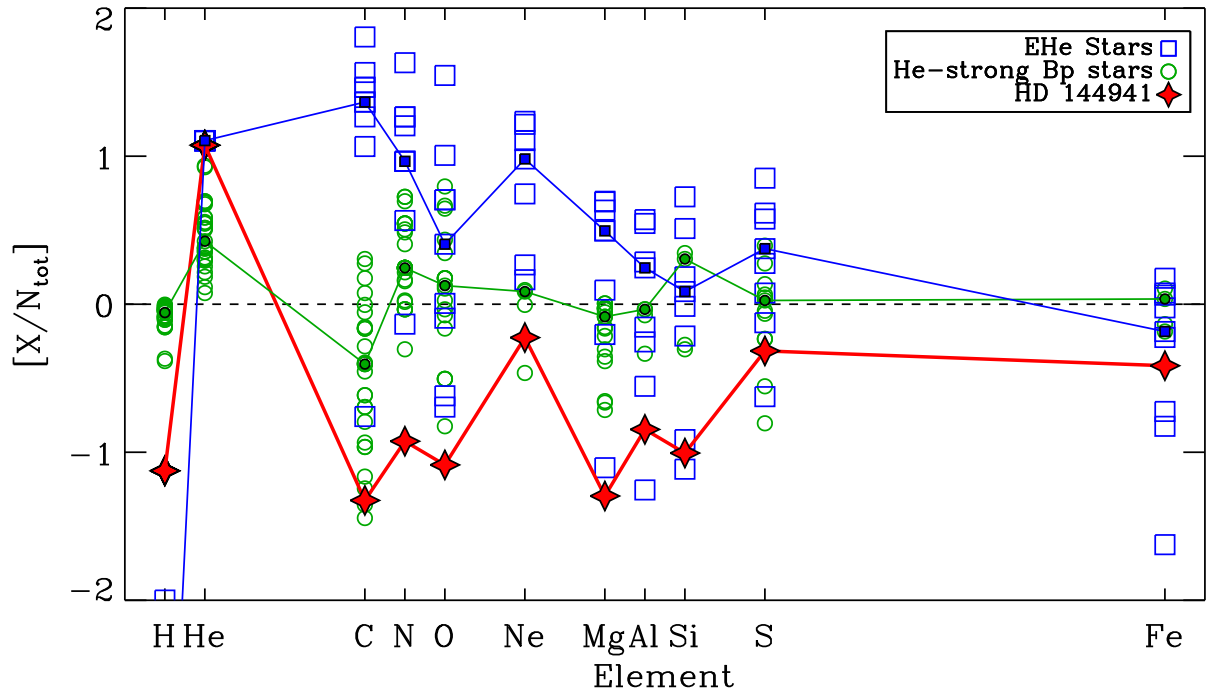}
\caption{Mean logarithmic surface abundances, as compared to solar, of EHe stars, He-s Bp stars, and HD 144941. The horizontal dashed line indicates solar abundances. Lines indicate median abundances for the two populations and for HD 144941. EHe stars exhibit larger departures from solar abundances than Bp stars for most elements, with generally greater enrichments as compared to solar. HD\,144941 by contrast is depleted in all elements except for He. HD\,144941 is within the observed range of variation of EHe stars for elements heavier than Mg. Its depletion in CNO and Ne is inconsistent with either population; however, its Ne abundance is enhanced with comparison to N and O, which is more typical of EHe stars than Bp stars.}
\label{ehe_bp_abundance_comparison}
\end{figure*}

Fig.\ \ref{ehe_bp_abundance_comparison} shows a comparison of the mean surface abundances determined for HD\,144941 to those of both EHe stars \citep{1999A&A...346..491J,2006ApJ...638..454P,2011ApJ...727..122P,2017ApJ...847..127P} and He-s mCP stars \citep{1997AA...324..949Z,1999A&A...345..244Z,2016A&A...587A...7P,2017MNRAS.467..437G,2017A&A...597L...6C,2019A&A...626A..94G,2019MNRAS.487.5922G}. In order to provide a common comparison between the two different populations, abundances are shown relative to the total number fraction $N_{\rm tot}$ in the solar atmosphere \citep{asplund2009}.

It is immediately apparent that HD\,144941's He abundance, while somewhat lower than that of the majority of EHe stars (generally 99\% or higher), is higher than the range of variation observed in mCP stars. Indeed, the He-s star with the highest surface He abundance, HD\,96446, has a number fraction of about 60\% \citep{2019A&A...626A..94G}, considerably lower than HD 144941's 95\% \citep{2017ApJ...847..127P}, and the majority of He-s stars have He fractions around 30\%.

Regarding metallic elements, HD 144941 is depleted in every element as compared to solar abundances. This contrasts to both the EHe stars, which are generally enriched in all elements except Fe, and He-s stars, for which the median abundances depart only slightly from solar values. HD\,144941 is furthermore on the low end of the range of variation observed for EHe stars, with the exception of N for which its abundance is about 1 dex below the most N-depleted star in this class. With one exception, all of the EHe stars are enhanced above solar in N and Ne, with abundances for heavier metallic elements being scattered both above and below solar metallicity. As compared to other EHe stars, HD\,144941 is an outlier mainly in its C and N abundances. 

He-s mCP stars, by contrast, generally have metallic abundances closer to solar values. N and Ne are also both slightly enhanced on average, therefore HD 144941 does not follow the pattern of mCP stars for these elements. Indeed, HD 144941 is outside of the range of variation measured for mCP stars for elements but C, Ne, and S. 

We conclude that HD\,144941's abundances do not closely match those of either population. However, it may be significant that HD 144941's Ne abundance is strongly enhanced as compared to both O and Mg, which is a typical pattern in EHe stars but not in He-s stars. 

\subsection{Magnetosphere}\label{magnetosphere}

\begin{figure*}
\includegraphics[width=\textwidth]{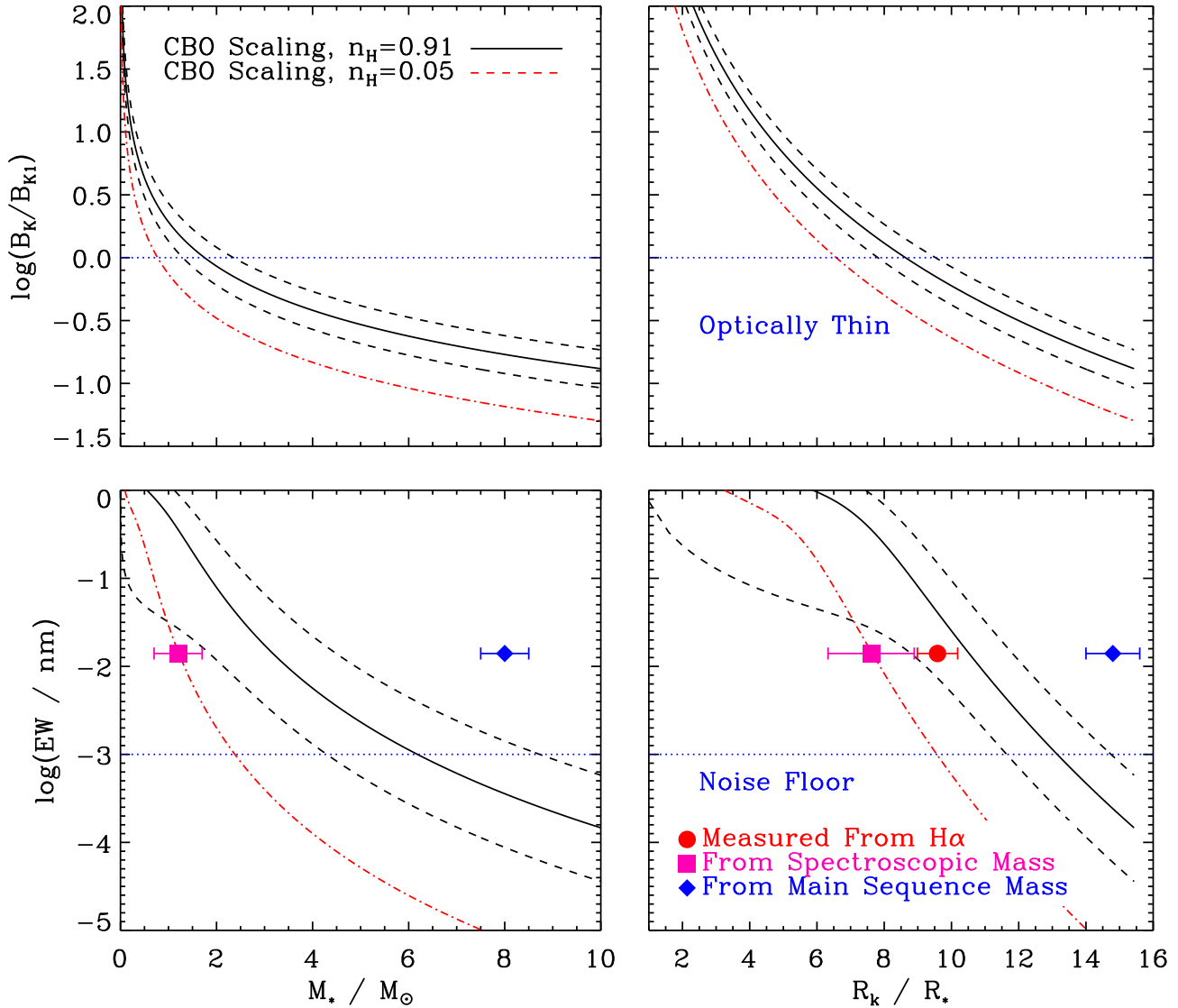}
\caption{Top: ratio of the magnetic field strength at the Kepler radius $B_{\rm K}$ to the value $B_{\rm K1}$ required for H$\alpha$ to be optically thick, as a function of mass $M_*$ (left) and Kepler radius $R_{\rm K}$ (right). Dashed curves indicate maximal uncertainties. An optically thick CM can be obtained only for a mass below the MS value. Bottom: predicted emission strength as a function of $M_*$ and $R_{\rm K}$. A low $M_*$ and small $R_{\rm K}$ matches the observed emission strength, whereas a mass consistent with the MS value would yield an emission strength below the noise floor. Dot-dashed red lines indicate the predicted emission properties following adjustment for HD 144941's surface H abundance.}
\label{hd144941_rk_em}
\end{figure*}

A stellar wind is magnetically confined if the dimensionless wind magnetic confinement parameter \citep{ud2002} $\eta_* > 1$. This parameter gives the ratio of magnetic energy density to wind kinetic energy density at the stellar equator. Since the spherical symmetry of the wind is broken by magnetic confinement, mass-loss rates and wind terminal velocities measured using spherically symmetric models are not reliable \citep{petit2013}. To determine the mass-loss rate we adopted an analytic model developed for subdwarf stars\cite{2016A&A...593A.101K}, using the effective temperature and luminosity determined for HD\,144941, and metallicity $Z/{\rm Z_\odot} = 0.001$, obtaining $\log{\dot{M} / ({\rm M}_\odot~{\rm yr}^{-1})} = -10.6 \pm 0.1$ and $v_\infty \sim 260$~\kms. Using these wind parameters, the lower limit of 17 kG on the dipolar surface magnetic field inferred from the magnetic modulus, and the radius, we calculate $\log{\eta_*} > 8$. From $\eta*$ the Alfv\'en radius \citep[corresponding to the distance from the star at which wind ram pressure and magnetic pressure equalize;][]{ud2002} is $R_{\rm A} > 100~R_*$. 

One of the ESPaDOnS observations shows two distinct emission bumps in H$\alpha$ (Fig. \ref{specplot}). The emission pattern is that typically seen in the Centrifugal Magnetospheres (CMs) of magnetic early B-type stars \citep{petit2013}. This double-humped emission pattern is superficially similar to the shell emission seen in classical Be stars viewed with the disk normal approximately perpendicular to the line of sight; the key difference is that whereas the shell emission of a classical Be star is contained inside $\pm$\vsini, CM emission occurs at velocities greater than $\pm$\vsini, as is this case for HD\,144941 (as indicated by the vertical dotted lines in Fig.\ \ref{specplot}). This general pattern has been observed in about 20 stars \citep{2020MNRAS.499.5379S}. In general these stars share 3 properties: they are rapid rotators ($P_{\rm rot} < 2$~d), they have strong magnetic fields (at least a few kG), and they are very young \citep{2019MNRAS.490..274S}. While HD\,144941 has a very strong magnetic field, it is an interesting exception as it is a slow rotator. Indeed, it is the only star with a rotational period in excess of about 2.7 d exhibiting CM-type emission, and the majority of CM stars have $P_{\rm rot} \sim 1$~d \citep{2019MNRAS.490..274S}. 

CMs are formed due to a combination of magnetic and centrifugal effects, as described by the Rigidly Rotating Magnetosphere \citep[RRM;][]{town2005c} model. As in any stellar magnetosphere, the magnetically trapped wind plasma is forced into corotation with the star by the photospheric magnetic field. This corotation leads to centrifugal force on the plasma. If the centrifugal force is negligible, magnetically confined plasma will collect in the magnetic equatorial plane, and then fall back to the star under the influence of gravity. If the centrifugal force is stronger than gravity, infall is prevented, and the plasma accumulates until being ejected outwards in centrifugally driven magnetic reconnection \citep{ud2008, 2020MNRAS.499.5379S, 2020MNRAS.499.5366O}. In the RRM model, plasma collects in an accumulation surface defined by the minima of the gravitocentrifugal potential along each magnetic field line, with an inner edge at the Kepler corotation radius $R_{\rm K}$, defined as the point at which gravitational and centrifugal forces balance \citep{town2005c}. In general, for a tilted dipole, this leads to the collection of plasma in a warped disk, with the densest parts of the CM at the intersections of the magnetic and rotational equatorial planes \citep{town2005c}. Since there is no plasma below $R_{\rm K}$, there is no emission at velocities below $(R_{\rm K}/R_*)v\sin{i}$, leading to the appearance of two emission bumps due to line formation in the two CM clouds \citep{town2005c,petit2013,2020MNRAS.499.5379S}. 

To estimate $R_{\rm K}$ directly from H$\alpha$, we determined the innermost wavelengths of the emission bumps by comparing the ESPaDOnS observation with emission to the one without, finding 655.95 nm and 656.42 nm for the blue and red bumps respectively. In the rest frame of the star, with a radial velocity of $-45$~\kms, these wavelengths correspond to projected velocities of $-105$~\kms~and $+109$~\kms. Due to the corotation of the magnetically confined plasma with the star, there is a linear relationship between projected velocity $v$ and projected distance $r$ from the star: $v/v\sin{i} = r/R_*$. Since \vsini~$= 11.2 \pm 0.7$~\kms, these wavelengths therefore correspond to a distance of about $9.6 \pm 0.7 R_*$, which we take to be $R_{\rm K}$, where the uncertainty is propagated from the uncertainty in \vsini. Notably, this Kepler radius is about 1/10 of the Alfv\'en radius. Corotation out to several times $R_{\rm K}$ is therefore very likely. 

Due to the close relationship between centrifugal force, gravity, and the existence of a CM, HD 144941's CM emission can in principle be used as an independent estimator of the star's mass. The Kepler radius is related to the dimensionless critical rotational parameter \citep{ud2008} $W$ as $R_{\rm K} = W^{-2/3}$, where $W = v_{\rm eq}/v_{\rm orb}$. The equatorial rotational velocity $v_{\rm eq} = 12.4 \pm 1$~\kms~is obtained directly from $R_*$ and $P_{\rm rot}$. The orbital velocity $v_{\rm orb}$, or the velocity required to maintain a Keplerian orbit at the stellar surface, is given by \citep{ud2008} $v_{\rm orb} =  (GM_*/R_*)^{1/2}$, where G is the gravitational constant. 

Adopting the spectroscopic mass of $1.2 \pm 0.5$~\msun~gives $R_{\rm K} = 7.6 \pm 1.3~R_*$, consistent within uncertainties with the value inferred from H$\alpha$. Conversely, if we use $M_* = 8.0 \pm 0.5$~\msun~as would be inferred from MS evolutionary models \citep{ekstrom2012} and the star's position on the HRD (assuming it to be a misidentified MS star), the Kepler radius is instead $14.8 \pm 0.8~R_*$. For MS stars, there is an excellent correlation between the value of \rk~measured from H$\alpha$ and the value calculated from first principles \citep{2020MNRAS.499.5379S}, although in most cases the measured value is about 25\% {\em larger} than the theoretical value: while the reasons for this offset are unclear, a similar difference is also seen in the case of HD\,144941. Importantly, there is no case for which the theoretical Kepler radius is larger than the Kepler radius measured from H$\alpha$ (as would need to be the case if HD\,144941 is really a MS star).

An important caveat in this analysis is that there are only four H$\alpha$ observations, and emission is seen in only one. Since coverage of the rotational phase curve is far from complete, it is by no means certain that $R_{\rm K}$ has been securely identified. Instead, the value from H$\alpha$ is a lower limit. CM emission exhibits an approximately sinusoidal variation across the line due to the rotational modulation of the corotating CM, with the strongest emission typically coinciding with the maximum projected distance from the star. This was first noted in the spectrum of $\sigma$ Ori E by \cite{walborn1974}; \cite{town2005a} demonstrated that this variability is a direct consequence of the plasma distribution predicted by the RRM model, an analysis extended by \cite{2015MNRAS.451.2015O} using an `arbitrary' RRM model to account for a magnetospheric structure extrapolated from a surface magnetic field inferred from Zeeman Doppler Imaging \citep{pk2002}. Similar variations have since been noted in a number of CM stars \citep[e.g.][]{leone2010,bohl2011,grun2012,rivi2013,2015MNRAS.451.1928S,2016MNRAS.460.1811S,2017MNRAS.465.2517W,2018MNRAS.475..839S,2020MNRAS.499.5379S}. It is therefore only at this phase of peak emission that $R_{\rm K}$ can be securely identified \citep{2020MNRAS.499.5379S}. Therefore it cannot be ruled out on the basis of this observation alone that $R_{\rm K}$ might be large enough to be compatible with the value inferred from the MS mass. This analysis is also sensitive to the value of \vsini: if \vsini~$\sim 7$~\kms, then $R_{\rm K}$ would indeed be about $14.8~R_*$ using the inner emission bump velocities determined here. We note, however, that \vsini~was measured from the C~{\sc II} doublets in the wings of H$\alpha$ together with the Zeeman splitting; since significant turbulent broadening is not expected \citep{2019MNRAS.487.3904M,2020ApJ...900..113J} or observed \citep{2018MNRAS.475.5144S,2013MNRAS.433.2497S} for stars with extremely strong photospheric magnetic fields, and HD 144941 is not a pulsator, there should not be significant additional line broadening mechanisms that have not been accounted for (although precise reconstruction of the detailed surface magnetic field may introduce some additional broadening for the Zeeman $\sigma$ components). Furthermore, a higher $R_{\rm K}$ seems unlikely in the context of the emission strength, as explored in the following.

Because H$\alpha$ emission is formed within an optically thick CM, the emission strength scales with the area of the emitting region \citep{2020MNRAS.499.5379S,2020MNRAS.499.5366O}. The optically thick area is itself essentially a function of the strength of the equatorial magnetic field $B_{\rm eq} = 0.5 B_{\rm d}$ (where $B_{\rm d}$ is the surface strength of the magnetic dipole) at the Kepler radius, given by $B_{\rm K} = B_{\rm eq}/R_{\rm K}^3$. This quantity regulates the centrifugal breakout (CBO) density at $R_{\rm K}$ \citep{2020MNRAS.499.5379S, 2020MNRAS.499.5366O}. In the CBO process, plasma accumulates in the CM up to the point at which the gas pressure exceeds the magnetic pressure, past which the magnetic field is no longer able to confine the plasma and it is ejected away from the star via centrifugally driven reconnection \citep{town2005c,ud2008}. In contrast to naive expectations from two-dimensional magnetohydrodynamic simulations that this should result in an emptying of the entire magnetosphere \citep{ud2008}, CBO events instead seem to occur continuously on small spatial scales, maintaining the CM in an approximately steady state \citep{2020MNRAS.499.5379S}.

The CBO density scaling leads directly to a scaling relationship for emission strength with $B_{\rm K}$ \citep{2020MNRAS.499.5366O}, which holds for essentially all known H$\alpha$-bright CM host stars \citep{2020MNRAS.499.5379S,2021arXiv210309670S,2021MNRAS.tmp..908S}. There is therefore every reason to expect that it holds in the case of HD\,144941 as well. As a consequence, if both $R_{\rm K}$ and $B_{\rm eq}$ are constrained, since $R_{\rm K}$ depends on $M_*$ the emission strength can also be used as an independent estimator of the consistency of $R_{\rm K}$ and $M_*$. 

Emission strength was measured from the H$\alpha$ EWs inside the highlighted regions in Fig.\ \ref{specplot}, subtracting the EW of the observation with emission from the observation without. The combined EW of the two bumps is $0.014 \pm 0.001$~nm. This is the weakest emission ever measured from a CM \citep{2020MNRAS.499.5379S}. 

We used the analytic CBO model \citep{2020MNRAS.499.5366O} to calculate the expected emission EW, over a mass range of $0.1 {\rm M_\odot} < M_* < 10 {\rm M_\odot}$. $P_{\rm rot}$, \teff, and $R_*$ were fixed, and we used $B_{\rm d} = 22^{+5}_{-3}$~kG. The range of $B_{\rm d}$ was chosen based on the maximum magnetic modulus of 17 kG and the median, maximum, and minimum ratios between the measured magnetic modulus and the inferred surface dipole strength in a large sample of mCP stars \citep{landmat2000}. We assumed a nominal projection angle of the disk normal to the line of sight of $45^\circ$, with minimal and maximal values of $0^\circ$ and $89^\circ$ (while the inclination angle of the rotational axis is constrained, magnetic data are insufficient to constrain the tilt angle of the magnetic axis from the rotational axis). The ratio of the source function of disk to star was fixed at 0.75, as appropriate for an isothermal disk with an \teff~about half that of the star, and as is generally a good match to observations \citep{2020MNRAS.499.5366O}. Since HD\,144941's CM is significantly further from the star than that of most H$\alpha$-bright CM stars, this assumed source function ratio may not be appropriate; since the CM would presumably be cooler than if it were closer to the star, the source function ratio would be lower and the emission consequently weaker for a given surface magnetic field strength. Note however that even for the maximum ratio of unity the results are qualitatively unchanged. 

The ratio $\log{(B_{\rm K}/B_{\rm K,1})}$ is shown in the top panels of Fig. \ref{hd144941_rk_em} as a function of $M_*$ and $R_{\rm K}$, where $B_{\rm K,1}$ is the strength of the magnetic field required to obtain an optical depth of 1 at the Kepler radius. The CM plasma is only optically thick when $\log{(B_{\rm K}/B_{\rm K,1})} > 0$. As can be seen, the maximum Kepler radius which can support an optically thick disk is $R_{\rm K} = 9~R_*$, much smaller than the value that would be inferred from the MS mass, but consistent with the values measured directly from H$\alpha$ and inferred from the spectroscopic mass. In fact, optically thick emission can be obtained only if $M_* < 2.5$~\msun. The bottom panels of Fig. \ref{hd144941_rk_em} show the predicted emission strength $\log{{\rm EW}}$ as a function of $M_*$ and $R_{\rm K}$. Emission strength consistent with the measured value is obtained for $M_* \sim 3$~\msun, and for a Kepler radius $R_{\rm K} \sim 10~R_*$. The large uncertainties (dashed curves in Fig. \ref{hd144941_rk_em}) are due to the conservative adoption of minimal and maximal values for the disk projection angle, which are highly uncertain since the tilt angle of the magnetic axis from the rotational axis is unknown, as well as due to the large uncertainties in $B_{\rm d}$. Notably, even with these maximal uncertainties, an 8~\msun~star with HD 144941's radius, surface magnetic field strength, and rotation period would have $B_{\rm K}$ far below the value necessary to maintain optically thick emission. Furthermore, the emission from such a star's CM would be at or below the noise floor determined from the uncertainty in the measured EW of the emission bumps. 

The surface magnetic field strength of HD\,144941 is only an estimate. In order for an 8~\msun~star with a 14~day rotation period to exhibit CM-type emission of the strength observed for HD\,144941, its surface magnetic field strength would need to be about 56~kG, over 3$\times$ the maximum value of $\langle B \rangle$, and far stronger than has been observed for any other magnetic OB star. Indeed, in this case HD\,144941 would have a significantly stronger surface magnetic field than the current record-holder, Babcock's Star, which has a maximum surface strength \citep{1997AstL...23..465K} of 44 kG, and a magnetic modulus \citep{1960ApJ...132..521B} of $\langle B \rangle_{\rm max} = 34$~kG.

These calculations have so far assumed a standard H mass fraction in the wind, as appropriate for the MS magnetic hot stars with which the CBO density scaling was calibrated. However, the much lower H abundance in the the photosphere of HD\,144941 as compared to the solar value suggests that H should also be depleted in the stellar wind and, therefore, in the magnetosphere. This means that it should be more difficult for HD\,144941's magnetosphere to become optically thick in H$\alpha$, since the magnetic field must confine a larger relative number of He and metallic ions in order to confine a given quantity of H ions. The dot-dashed red curves in Fig. \ref{hd144941_rk_em} show the result of adjusting the CBO scaling for the lower H abundance, under the assumption that the abundance in the wind matches that of the photosphere \citep[following Eqn. A4 in the appendix of][]{2020MNRAS.499.5366O}. This adjustment results in an almost perfect agreement with the emission properties inferred from the spectroscopic mass. Conversely, achieving the measured emission level with the mass inferred from MS evolutionary models would require a minimum surface magnetic field strength of 150 kG, well beyond that of Babcock's Star.

We conclude that an EHe star with a mass of around 1 \msun~would possess a Kepler radius and an emission strength closely matching the values directly measured from HD 144941's emission profile, while on the other hand a MS star with a mass of 8 \msun~would have too large of a Kepler radius and, unless its surface magnetic field strength is unreasonably strong compared to both the available measurements and all known MS magnetic hot stars, no detectable emission. 

\subsection{Summary}

Several lines of evidence independently suggest that HD 144941 is in fact an EHe star. First (\S~\ref{abund}, Fig.\ \ref{ehe_bp_abundance_comparison}), its surface He abundance is much higher than is observed for He-s mCP stars, but consistent with the range of variation in EHe stars. Second (\S~\ref{abund}, Fig.\ \ref{ehe_bp_abundance_comparison}), while its depletion of metallic elements (i.e.\ elements heavier than He) relative to the Sun is somewhat peculiar in comparison to EHe stars, it is within the range of variation of this class, whereas He-s mCP stars exhibit systematically higher abundances. Third, (\S~\ref{stellarpar} and \ref{specmass}, Fig.\ \ref{hrd}), the $1.2 \pm 0.5$~\msun~spectroscopic mass inferred from its {\em Gaia} Early Data Release 3 parallax \citep{2020arXiv201206420T} and the surface gravity determined from non-Local Thermodynamic Equilibrium spectroscopic modelling \citep{2017ApJ...847..127P} is much less than can be consistent with a MS star of HD\,144941's effective temperature and luminosity. Fourth (\S~\ref{magnetosphere}, Fig.\ \ref{hd144941_rk_em}), the emission seen in H$\alpha$ (Fig.\ \ref{specplot}) is consistent with an origin in a CM \citep{petit2013}; given the very slow rotation of this star in comparison with MS stars showing CM-type emission, all of which are rapid rotators \citep{2019MNRAS.485.1508S}, such emission can only be produced if the stellar mass is below about 2~\msun \citep{2020MNRAS.499.5379S,2020MNRAS.499.5366O}. 

Surface abundances alone are not enough to confidently assign HD\,144941 to the EHe class, since magnetic fields produce surface abundance anomalies that mask the underlying bulk abundances of the star. By the same token, the discrepancies between HD\,144941's metallic abundances and those of `normal' EHe stars are not sufficient to reject the star from the class. The most convincing evidence that HD\,144941 is not a MS star comes from the mass inferred spectroscopically and from modelling its magnetospheric H$\alpha$ emission, which in both cases yields a mass of around 1 \msun, which is much too low for the star to be on the MS. 

\section{Discussion}\label{sec:discussion}

The detection of such a strong magnetic field in an EHe star raises the question of whether EHe stars in general are magnetic, since after all they are all merger products. While high-resolution magnetic measurements are not available in the literature for other stars of this class, there is reason to doubt that magnetic fields are ubiquitous in this class. No other EHe star is known to exhibit photometric rotational modulation \citep{2018MNRAS.475L.122J,2020MNRAS.495L.135J}. Furthermore, HD\,144941 stands out as a chemically peculiar EHe star. Its low Fe abundance is believed to originate with the low metallicities of its progenitors \citep{2017ApJ...847..127P}. However, in contrast to other EHe stars, it has reduced C, N, O, and Ne abundances, which are expected to be enhanced due to nuclear burning in the merger process \citep{2017ApJ...847..127P}. 

Assuming HD\,144941's peculiar surface abundances to be informed by its formation history as well as by the presence of a strong magnetic field, it is reasonable to speculate that the evolutionary history of HD\,144941 differs from that of other EHe stars. There are two known pathways that can lead to the formation of a subdwarf OB star: stripping by a binary companion, and the merger of a WD binary system. There is no evidence that HD\,144941 is a double star. Its radial velocities are absolutely stable over the 14-year span separating the FEROS and ESPaDOnS spectra, nor are there any anomalies in the line profiles or in the spectrum that would suggest the presence of such a star. Since the stripped star will necessarily be the donor in the interaction, the companion should be more massive, at least as bright, and therefore more easily detected than the stripped star. Since no such companion is detectable it is highly unlikely that HD\,144941 is a stripped star, leaving only a merger as a viable formation pathway.

Assuming HD\,144941 to be a merger remnant, we are left with the question of its peculiar surface abundances, which do not reflect the CNO enhancement expected from nuclear burning in the merger process of two He-WD stars. Furthermore, if both progenitors were WDs, the inferred mass of about 1 \msun~would require at least one of them to have been a CO or ONe WD, making the subsolar CNO abundances all the more puzzling. One possibility is that entropy sorting may have resulted in the heavier elements from the more massive star having sunk to the core \citep[e.g.][]{2008A&A...488.1017G}. Another possibility may be that HD\,144941 is not the product of two He-WDs, but a He-WD+MS merger; indeed, several of the star's properties, e.g.\ its luminosity and its relatively low He abundance as compared to other EHe stars, are consistent with this scenario \citep[e.g.][]{2017ApJ...835..242Z}, which further predicts that CNO products should rapidly sink following the merger process. A final possibility, as discussed in \S~\ref{specmass}, is that magnetic pressure may have increased the pressure broadening of the Balmer wings, leading to an erroneously high measurement of the surface gravity and, hence, the spectroscopic mass.

Two merger scenarios have been explored for the formation of EHe stars: slow mergers, and fast mergers \citep{2012MNRAS.419..452Z}. In slow mergers the lower-mass star breaks apart into a disk, which accretes onto the primary over several Myr. In fast mergers, accretion is complete within a few minutes. In the case of a fast merger the secondary forms a hot coronal envelope surrounding the primary, which then forms a thick convection zone as it accretes. Interestingly, the surface abundance patterns of the majority of EHe stars are consistent with slow mergers \citep{2011ApJ...727..122P}, whereas HD\,144941's abundances have been suggested to be more compatible with either a fast merger or a composite merger i.e.\ a combination of the two processes \citep{2017ApJ...847..127P}. The large, energetic convection zone provided by a fast merger is precisely the physical environment within which a magnetic dynamo can be sustained \citep{2019Natur.574..211S}. Thus, while the detection of HD\,144941's powerful magnetic field is clear confirmation that binary mergers provide a pathway to the formation of fossil fields, the absence of indications for magnetic fields in other EHe stars, and the possibility that HD\,144941 formed via a distinct merger process as compared to other stars of this class, suggest that the type of merger -- fast or slow -- is a key factor in determining whether or not a fossil magnetic field will be produced. 

EHe stars are short-lived objects which are expected to rapidly evolve onto the WD cooling sequence. In this case, assuming that the unsigned magnetic flux $\Phi = BR_*^2$ is conserved, the surface magnetic field should intensify by a factor of $(R_i/F_f)^2$, where $R_i$ and $R_f$ are the initial and final radii. Taking a typical WD radius as around $0.01~{\rm R_\odot}$ and a present surface magnetic field of about 20 kG, flux conservation would imply that HD\,144941 will become a He WD with a surface magnetic field of about $10^9$ G. This is comparable to the upper limit of the magnetic field strengths observed in magnetic WD (MWDs) \citep[e.g.][]{2020AdSpR..66.1025F}. Since HD\,144941 is apparently a merger product, this supports the scenario in which at least some MWDs -- and in particular those with extremely strong magnetic fields -- are produced in WD merger events \citep{2020AdSpR..66.1025F}.

The alternatives to the merger origin is that MWDs are the descendents of magnetic Ap/Bp stars, with surface fields again reflecting flux conservation from the MS, or that they are a result of fossil fields produced in the convective core dynamos of MS stars. The latter scenario is supported by asteroseismic irregularities in red giant stars indicative of strong core magnetic fields \citep{2015Sci...350..423F,2016Natur.529..364S}. The former scenario is challenged by the overall 20\% incidence of MWDs in the local 20 pc volume \citep{2019A&A...628A...1L}, about twice the incidence of strong magnetic fields amongst MS OBA stars \citep{2017MNRAS.465.2432G,2019MNRAS.483.2300S}. Furthermore, a similar flux conservation mechanism has been proposed as the origin for magnetar B fields, however a careful comparison of the properties of the MS population of magnetic early B-type stars with the neutron star population seems to rule this scenario out \citep{2021MNRAS.504.5813M}, in turn making simple flux conservation from the MS less likely as an origin for strong-field MWD stars. Since flux conservation from the MS seems increasingly unlikely, we are left with fossilized core dynamos or fossilized merger dynamos. Indeed, both scenarios may play a role in the origin of MWDs: for instance, mergers may result in strong-field MWDs ($10^6-10^9$ G), while core dynamos might leave behind the weak-field stars (down to $10^3$ G). 

Recently, \cite{2021arXiv210603363V} reported the discovery of an apparent post-merger (Asymptotic giant branch+He-WD) subdwarf B-type star which demonstrates indications in its spectrum of a CM and of Zeeman splitting indicative of surface magnetic field of several hundred MG. The light-curve of this object is very similar to that of HD\,144941, being both perfectly periodic and highly structured, although much shorter at only about 0.07~d due to the object's much smaller radius. This object may be a lower-mass analogue to HD\,144941, observed closer to the merger event. 

Mergers have of course also been suggested as the origins of magnetic fields amongst MS hot stars \citep{2019Natur.574..211S}. At present however there are only two candidate magnetic B stars that have been proposed as merger products, HR\,2949 \citep{2015MNRAS.449.3945S} and $\tau$ Sco \citep{2019Natur.574..211S}. In both cases the primary evidence for the stars being merger products are discrepancies in the ages of the stars, which can be resolved if they have been rejuvenated by stellar mergers \citep{2016MNRAS.457.2355S}. In the case of $\tau$ Sco, the most thoroughly studied of the two, the merger hypothesis is not yet confirmed, as the irregularities in its present age may simply be a consequence of magnetic effects on stellar evolution \citep{2021MNRAS.504.2474K}.

\section{Conclusions}\label{sec:conclusions}

We have detected an extremely strong magnetic field field (\bz~$\sim -8$~kG, \bmodulus~$\sim$~17 kG) in high-resolution ESPaDOnS spectropolarimetry of HD\,144941. 

The previously reported 13.9~d rotation period has been refined via comparison of the {\em K2} and {\em TESS} light curves, and this period accurately phases ESPaDOnS and FEROS observations acquired 15 years apart. The accurate phasing of the ESPaDOnS and FEROS datasets indicates that the photometric modulation is indeed a conseqeuence of rotation, as is the line profile variability. 

HD\,144941's surface abundances are a poor match to those of other EHe stars: it has a higher H abundance, a lower He abundance, and is extremely metal-poor, with no indication of nuclear burning products in its CNO and Ne abundances. On the other hand, its He abundance is much higher than that of any other MS He-s star, and its low metallic abundances are not reflective of this population either. Future examination of this object should attempt to determine whether its surface abundance measurements have been affected by the star's strong magnetic field, and whether they are truly reflective of the star's bulk composition or (as is usually the case with mCP stars) entirely a surface phenomenon. 

HD\,144941 is the most slowly rotating star to display H$\alpha$ emission consistent with an origin in a centrifugal magnetosphere. Given the slow rotation and lower limit on the surface magnetic field strength, for this emission to be detectable, HD\,144941 must have a mass of around 1 \msun, consistent with its being a WD merger product rather than a mis-identified He-s star. This mass is consistent with the star's spectroscopic mass, although the conclusion is dependent upon the surface gravity determined via spectroscopic analysis, which varies between studies. Future spectroscopic analysis should attempt to determine to what degree the star's surface gravity measurements have been affected by magnetic pressure broadening. 

Assuming that HD\,144941 is in fact an EHe star, the detection of a strong magnetic field is strong evidence that binary mergers can produce fossil magnetism. However, the lack of evidence for associated magnetic phenomena (in particular, rotational modulation) in other EHe stars argues that mergers do not always produce magnetic fields. HD\,144941's chemical peculiarity in comparison to other EHe stars may also point to a different origin scenario for HD\,144941. Since there is no evidence that HD\,144941 is a binary, it is unlikely to be a stripped star and, if truly a subdwarf, is most likely a merger product. It has been suggested that HD\,144941's abundances may reflect a rapid merger, whereas other EHe stars formed via slow mergers. In the future it will be necessary to determine whether or not EHe stars are a generally magnetic class. In the case that they are not, this would point to a sensitive dependence on the type of merger -- fast or slow -- for the ability of a merger-powered magnetic dynamo of sufficient intensity to leave behind a detectable fossil to be generated. 

\section*{Acknowledgements}
The authors warmly acknowledge the thoughtful review provided by Dr.\ Richard Townsend, which greatly improved the quality of this paper. This work is based on observations obtained at the Canada-France-Hawaii Telescope (CFHT) which is operated by the National Research Council of Canada, the Institut National des Sciences de l'Univers (INSU) of the Centre National de la Recherche Scientifique (CNRS) of France, and the University of Hawaii; and at the La Silla Observatory, ESO Chile with the MPA 2.2 m telescope. Some of the data presented in this paper were obtained from the Mikulski Archive for Space Telescopes (MAST). This paper includes data collected by the {\em TESS} mission, which are publicly available from the Mikulski Archive for Space Telescopes (MAST). Funding for the {\em TESS} mission is provided by NASA’s Science Mission directorate. This work has made use of the VALD database, operated at Uppsala University, the Institute of Astronomy RAS in Moscow, and the University of Vienna. M.E.S. acknowledges financial support from the Annie Jump Cannon Fellowship, supported by the University of Delaware and endowed by the Mount Cuba Astronomical Observatory. O.K. acknowledges support by the Swedish Research Council and the Swedish National Space Board. J.L-B. acknowledges support from FAPESP (grant 2017/23731-1). The material is based upon work supported by NASA under award number 80GSFC21M0002.

\section*{Data Availability Statement} 
Reduced ESPaDOnS spectra are available at the CFHT archive maintained by the CADC\footnote{\url{https://www.cadc-ccda.hia-iha.nrc-cnrc.gc.ca/en/}}. Reduced FEROS spectra are available at the ESO archive\footnote{\url{http://archive.eso.org/eso/eso_archive_main.html}}. {\em Kepler-2} and {\em TESS} data are available at the MAST archive\footnote{\url{https://mast.stsci.edu/portal/Mashup/Clients/Mast/Portal.html}}. Data in all archives can be found via standard stellar designations. 

\bibliography{bib_dat}{}

\end{document}